\documentclass[11pt,english]{article}
\usepackage[T1]{fontenc}
\usepackage[latin1]{inputenc}
\usepackage{graphicx}
\usepackage{amssymb}

\makeatletter

\providecommand{\LyX}{L\kern-.1667em\lower.25em\hbox{Y}\kern-.125emX\@}

 \newcommand{\lyxaddress}[1]{
   \par {\raggedright #1 
   \vspace{1.4em}
   \noindent\par}
 }

\usepackage{babel}
\makeatother
\begin{document}

\title{The theoretical model of a magnetic white dwarf. }

\author{I. Bednarek, A. Brzezina, R. Ma\'{n}ka, M. Zastawny-Kubica}

\maketitle

\lyxaddress{Department of Astrophysics and Cosmology, Institute of Physics, University
of Silesia, Uniwersytecka 4, 40-007 Katowice, Poland. }

\begin{abstract}
\emph{In this paper a theoretical model of a magnetic white dwarf
is studied. All numerical calculations are performed under the assumption
of a spherically symmetric star. The obtained equation of state is
stiffer for the increasing value of magnetic field \textbackslash{}(B\textbackslash{}).
The numerical values of the maximum mass and radius are presented.
Sensibility of the results to the choice of the magnetic field strength
is evident. Finally the departure from the condition of isothermality
of degenerated electron gas in gravitational field is discussed.}
\end{abstract}

\section{Introduction}

The possibility of the existence of white dwarfs with large magnetic
fields was predicted by Blackett in 1947. Ever since Kemp et al. in
their paper presented results of the first detection of magnetic white
dwarfs, these kind of stars have been intensively studied (Kemp et
al. 1970; Putney 1995; Reimers et al. 1996; Jordan 1992; Chanmugan
1992; Ostriker et al. 1968; Mathews et al. 2000). However, only a
small minority (the exact percentage remains uncertain) of white dwarfs
have detectable magnetic fields. The study of magnetic properties
of white dwarfs are expected to provide information on the role played
by magnetic fields in the stellar formation and the main sequence
evolution. Magnetic fields in white dwarfs are interpreted as the
relic of magnetic fields of their progenitors. Thus the evolution
of the main sequence stars with fields of only a few Gauss leads to
white dwarfs with weak magnetic fields which remain below the threshold
detection. Magnetic Ap and Bp stars are considered themselves as progenitors
of magnetic white dwarfs with the field strength $>100\, \, MG$.
\\
 The aim of this paper is to show how the assumption of finite temperature
and nonzero magnetic field influences white dwarf parameters. In the
presence of magnetic field the magnetic pressure which corresponds
to the Lorence force has to be taken into account. However, its inclusion
causes the deformation of a star from spherical configuration, (Konno
et al. 1999). In this paper all calculations are performed with the
assumption that the averaged value of the energy momentum tensor gives
us the condition for small deformation of a star and allows to treat
a star as a spherically symmetric one. This approximation is connected
with the existence of the critical density $\rho _{\mathrm{crit}}$
below which a star became deformed. The gained results allow to construct
the mass-radius relation for a white dwarf with the magnetic field
and at finite temperature which has been chosen to be equal $10^{7}$
$K$. The description of the interior of a white dwarf has been made
in terms of degenerate, uniform, isothermal, electron gas. Using the
equation of state of classical gas the atmosphere model has been formulated.
The atmosphere extent, luminosity, effective temperature and the value
of the critical density $\rho _{\mathrm{crit}}$ as functions of the
temperature, density and magnetic field strength have been examined.
The determination of the critical point allows to neglect the deformation
of the star and perform all calculations in the spherically symmetric
approximation. \\
This paper is organized as follows.\\
Section 2 outlines the considered model and contains assumptions which
are indispensable in determining the form of the energy momentum tensor
and the value of the critical density $\rho _{\mathrm{crit}}$. In
section 3 the numerical results and the discussion are presented.

\section{The theoretical model of white dwarfs in the magnetic fields }

This paper presents the theoretical model of a white dwarf in which
the main contribution to the pressure which supports the star against
collapse comes from ultra-relativistic electrons . The electrons are
supposed to be degenerate with arbitrary degree of relativity $x=\frac{k_{\mathrm{F}}}{m_{\mathrm{e}}}$
(in this paper the units in which $c=\hbar =1$ were used ). The value
of $x$ is a function of a density. Ions provide the mass of the star
but their contribution to the pressure is negligible. The complete
and more realistic description of a white dwarf requires taking into
consideration not only the interior region of the star but also the
atmosphere. The isothermal interior of a white dwarf is completely
degenerated and covered by non-degenerate surface layer which is in
radiative equilibrium. Thus the white dwarf matter consists of electrically
neutral plasma which comprises positively charged ions and electrons.\\
The Lagrangian density function in this model can be represented as
the sum \begin{equation}
{\mathcal{L}}={\mathcal{L}_{\mathrm{e}}}+{\mathcal{L}_{\mathrm{G}}}+{\mathcal{L}_{\mathrm{QED}}},\label{lagra}\end{equation}
 where ${\mathcal{L}_{\mathrm{e}}}$, ${\mathcal{L}_{\mathrm{G}}}$
describe the electron and gravitational terms, respectively. The ${\mathcal{L}_{\mathrm{QED}}}$
is the Lagrangian density function of the QED theory \begin{eqnarray*}
{\mathcal{L}_{\mathrm{QED}}}=-\frac{1}{4}F^{\mu \nu }F_{\mu \nu }. &  & 
\end{eqnarray*}
\\
The electron part of the Lagrangian is given by \[
{\mathcal{L}_{\mathrm{e}}}=i\overline{\psi _{\mathrm{e}}}\gamma ^{\mu }D_{\mu }\psi _{\mathrm{e}}-m_{\mathrm{e}}\overline{\psi _{\mathrm{e}}}\psi _{\mathrm{e}},\]
 where $D_{\mu }$ is the covariant derivative defined as$\, \, D_{\mu }=\partial _{\mu }-ieA_{\mu }\, $\thinspace and
$e$ is the electron charge. The Dirac equation for electrons obtained
from the Lagrangian function ${\mathcal{L}}$ has the form \[
(i\gamma ^{\mu }D_{\mu }-m_{\mathrm{e}})\psi _{e}=0.\]
 The vector potential is defined as $A_{\mu }=\{A_{\mathrm{0}}=0,\, A_{\mathrm{i}}\}$
where \[
A_{\mathrm{i}}=-\frac{1}{2}\varepsilon _{\mathrm{ilm}}x^{\mathrm{l}}B_{\mathrm{0}}^{\mathrm{m}}.\]
 The symmetry in which uniform magnetic field B lies along the z-axis
has been chosen \[
B_{\mathrm{0}}^{\mathrm{m}}=(0,0,B_{\mathrm{z}}).\]
 The energy momentum tensor is the sum of the perfect fluid part $T_{\mu \nu }^{\mathrm{matt}}$\begin{eqnarray*}
T_{\mu \nu }^{\mathrm{matt}}=(\varepsilon +p)u_{\mu }u_{\nu }+g_{\mu \nu }p, &  & 
\end{eqnarray*}
 and the electromagnetic one $T_{\mu \nu }^{\mathrm{B}}$\begin{eqnarray*}
T_{\mu \nu }^{\mathrm{B}}=F_{\mu }^{\lambda }F_{\nu \lambda }-\frac{1}{4}g_{\mu \nu }F_{\alpha \beta }F^{\alpha \beta } &  & 
\end{eqnarray*}
 and can be calculated taking the quantum statistical average \begin{eqnarray*}
\bar{T}_{\mu \nu }=<T_{\mu \nu }>=T_{\mu \nu }^{B}+T_{\mu \nu }^{matt}. &  & 
\end{eqnarray*}
 In the cartesian coordinates the electromagnetic part of the energy
momentum tensor in the flat space-time has the form 

\begin{equation}
T_{\mu \nu }^{\mathrm{B}}=\left(\begin{array}{cccc}
 \frac{1}{2}B^{2} & 0 & 0 & 0\\
 0 & \frac{1}{2}B^{2} & 0 & 0\\
 0 & 0 & \frac{1}{2}B^{2} & 0\\
 0 & 0 & 0 & -\frac{1}{2}B^{2}\end{array}
\right),\label{tensormetryczny}\end{equation}
 and presents the example of the anisotropic pressure. In general
the pressure of the electron gas in the magnetic field can be written
as the sum of two parts \begin{eqnarray*}
P=P_{\mathrm{e}}+P_{\mathrm{QED}} &  & 
\end{eqnarray*}
 where $P_{\mathrm{e}}$ is the isotropic part of the pressure, $P_{\mathrm{QED}}$
can reach the value $\pm \frac{1}{2}B^{\mathrm{2}}$ and causes that
the pressure became anisotropic. On the other hand $T_{{\mu }{\nu }}^{\mathrm{B}}$
for polar coordinates can be written as follows \begin{eqnarray*}
\left(\begin{array}{cccc}
 \frac{1}{2}B^{2} & 0 & 0 & 0\\
 0 & -\frac{1}{2}B^{2}\cos 2\theta  & \frac{1}{2}B^{2}r\sin 2\theta  & 0\\
 0 & \frac{1}{2}B^{2}r\sin 2\theta  & \frac{1}{2}B^{2}r^{2}\cos 2\theta  & 0\\
 0 & 0 & 0 & \frac{1}{2}B^{2}r^{2}\sin ^{2}\theta \end{array}
\right). &  & 
\end{eqnarray*}
 Averaging over all directions allows to obtain the isotropic form
of the energy momentum tensor relevant for the spherically symmetric
configuration \begin{eqnarray*}
\left(\begin{array}{cccc}
 \frac{1}{2}B^{2} & 0 & 0 & 0\\
 0 & \frac{1}{6}B^{2} & 0 & 0\\
 0 & 0 & \frac{1}{6}B^{2} & 0\\
 0 & 0 & 0 & \frac{1}{6}B^{2}\end{array}
\right). &  & 
\end{eqnarray*}
 The assumption of the positivity of the total pressure $P$ has been
made \begin{eqnarray*}
P_{\mathrm{e}}-\frac{1}{2}B^{\mathrm{2}}\geq 0. &  & 
\end{eqnarray*}
 Any negative contribution to the pressure reduces it, gravity can
not be compensated and this diminishes the radius of the star. The
described above averaging approximates a star as a spherically symmetric
object with the radius $R$ which is smaller than the equatorial radius
of the anisotropic star. The case of an anisotropic star is calculated
in details in the paper by Konno et al. (Konno et al. 1999). The condition
of the positivity of the total pressure establishes a limiting value
of the density noted $\rho _{\mathrm{crit}}$ or the critical point
$x_{\mathrm{crit}}$ and puts limit on the spherically symmetric approximation
(the deformation of the star is neglected). For densities smaller
than the critical one the total pressure became negative and there
is no gravity compensation in this direction.\\
 The equations describing masses and radii of white dwarfs are determined
by the proper form of the equation of state. The aim of this paper
is to calculate the equation of state for the theoretical model of
magnetic white dwarfs with the assumption of finite temperature. \\
Properties of an electron in the external magnetic field have been
studied, for example by Landau and Lifshitz in 1938 (Landau et al.
1938). In this paper the effects of the magnetic field on an equation
of state of a relativistic, degenerate electron gas is considered.
The motion of free electrons in the homogeneous magnetic field of
the strength $B$ in the direction perpendicular to the field is confined
by the oscillatory force determined by the field $B$ and is quantized
into Landau levels with the energy $nB$, $n=0,1,\ldots $. In the
case of non-relativistic electrons their energy spectrum is given
by the relation \begin{eqnarray*}
E_{\mathrm{n,p}_{z}}=n\omega _{\mathrm{c}}+\frac{p_{\mathrm{z}}^{2}}{2m_{\mathrm{e}}} &  & 
\end{eqnarray*}
 where $n=j+\frac{1}{2}+s_{\mathrm{z}}$, the cyclotron energy $\omega _{c}=eB/m_{e}$,
$p_{z}$ is the momentum along the magnetic field and can be treated
as continuous. For extremely high magnetic field the cyclotron energy
is comparable with the electron rest mass energy and this is the case
when electrons became relativistic. Introducing the idea of the critical
magnetic field strength $B_{\mathrm{c}}=m_{\mathrm{e}}^{2}/\mid e\mid $
which equals $4.414\times 10^{\mathrm{13}}$ $G$, it is easy to distinguish
between non-relativistic and relativistic cases, for $B\geq B_{\mathrm{c}}$
the relativistic dispersion relation for the electron in the magnetic
field is given by \begin{eqnarray}
E_{\mathrm{n,p}_{\mathrm{z}}}=\sqrt{p_{\mathrm{z}}^{\mathrm{2}}+m_{\mathrm{e}}^{\mathrm{2}}+2eB_{\mathrm{z}}n} &  & \label{eq1}
\end{eqnarray}
 where $n$ is the Landau level, $p_{z}$ is the electron momentum
along z-axis and $m_{e}$ is the rest mass of the electron. Along
the field direction a particle motion is free and quasi-one-dimensional
with the modified density of states given by a sum \begin{equation}
2\int \frac{d^{\mathrm{3}}{p}}{(2\pi )^{\mathrm{3}}}\rightarrow 2\sum _{n=0}^{\infty }[2-\delta _{\mathrm{n0}}]\int \frac{eB_{\mathrm{z}}}{(2\pi )^{\mathrm{2}}}dp_{z},\end{equation}
 $\delta _{\mathrm{n0}}$ denotes the Kronecker delta (Mathews et
al. 2000). The spin degeneracy equals 1 for the ground ($n=0$) Landau
level and 2 for $n\geq 1$. The redefined density of states makes
the distinctive difference between the magnetic and non-magnetic cases.
The equation (\ref{eq1}) implies that for $n=0$, $E_{\mathrm{0}}=\sqrt{p_{\mathrm{z}}^{\mathrm{2}}+m_{\mathrm{e}}^{\mathrm{2}}}$
whereas for $n\geq 1$ $E_{n}=\sqrt{p_{\mathrm{z}}^{\mathrm{2}}+m_{\mathrm{e}}^{\star \mathrm{2}}}$.
These relations indicate that the quantity $m_{e}^{\star }$ defined
as $m_{\mathrm{e}}^{\star \mathrm{2}}=m_{\mathrm{e}}^{\mathrm{2}}+2eB_{\mathrm{z}}n$
can be interpreted as the effective electron mass which is different
from the electron mass $m_{\mathrm{e}}$ for $n\geq 1$. The number
density of electrons at zero temperature is given by the following
relation \begin{eqnarray*}
n_{\mathrm{e}}=\sum _{\mathrm{n=0}}^{n=n_{\mathrm{max}}}[2-\delta _{\mathrm{n0}}]2eB_{{z}}p_{\mathrm{e}}^{F}. &  & 
\end{eqnarray*}
 The maximum Landau level is estimated from the condition $(p_{\mathrm{e}}^{\mathrm{F}})^{\mathrm{2}}\geq 0$.
One can define the critical magnetic density $\rho _{\mathrm{B}}$\begin{equation}
\rho _{\mathrm{B}}=0.802\
 Y_{\mathrm{e}}^{\mathrm{-1}}\gamma ^{\mathrm{3/2}}g\, \, cm^{-3}\label{eqrhob}\end{equation}
 ($\gamma =B_{{z}}/B_{{c}}$ means a dimensionless parameter which
expresses the magnetic field strength through the value of the critical
field $B_{{c}}$) for densities lower than $\rho _{\mathrm{B}}$ only
the ground Landau level is present.\\
Assuming the finite temperature of the system which affects the electron
motion in the external magnetic field the number density of electrons
can be expressed now as \begin{eqnarray}
 & n_{\mathrm{e}}=\frac{2m_{\mathrm{e}}^{\mathrm{3}}\gamma }{4\pi ^{\mathrm{2}}}\sum _{\mathrm{n=0}}^{\infty }[2-\delta _{\mathrm{n0}}]\times  & \\
 & (I_{\mathrm{0,0,+}}(z/t,1+2\gamma n)-I_{\mathrm{0,0,-}}(z/t,1+2\gamma n)) & \nonumber 
\end{eqnarray}
 where the Fermi integral \begin{equation}
I_{\lambda \mathrm{,}\eta \pm }(u,\alpha )=\int \frac{(\alpha +x^{2})^{\lambda \mathrm{/2}}x^{\eta }dx}{e^{\left(\sqrt{\alpha +x^{\mathrm{2}}}\mp u\right)}+1}\end{equation}
 has been used. The $\mu _{\star }$ and $T_{\star }$ denote the
chemical potential and temperature in the star interior and $z=\mu _{\star }/m_{\mathrm{e}}$,
$t=k_{\mathrm{B}}T_{\star }/m_{\mathrm{e}}$, $u=z/t$ (Lai 2000;
Bisnovatyi-Kagan 2000). Finite temperature together with decreasing
value of the strength of the magnetic field tend to smear out Landau
levels. The introduction of the concept of Fermi temperature \begin{equation}
T_{\mathrm{F}}=E_{\mathrm{F}}/k_{\mathrm{B}}=(m_{\mathrm{e}}/k_{\mathrm{B}})\epsilon _{\mathrm{F}}\qquad (\mathrm{for}\rho <\rho _{\mathrm{B}}),\label{eqtf}\end{equation}
 and the magnetic temperature \begin{eqnarray}
 & T_{\mathrm{B}}={\frac{\Delta E_{\mathrm{B}}}{k_{\mathrm{B}}}} & \\
 & ={\frac{m_{\mathrm{e}}}{k_{\mathrm{B}}}}\left(\sqrt{1+2n_{\mathrm{max}}\gamma +2\gamma }-\sqrt{1+2n_{\mathrm{max}}\gamma }\right) & \label{eqtb}\nonumber 
\end{eqnarray}
 where $\Delta E_{\mathrm{B}}$ is the energy difference between the
$n_{\mathrm{L}}=n_{\mathrm{max}}$ level and the $n_{L}=n_{\mathrm{max}}+1$
level, is the convenient way for describing how the properties of
free electron gas change at finite temperature and at the presence
of the magnetic field. For $\rho =\rho _{\mathrm{B}}$ these temperatures
equal $T_{\mathrm{F}}=T_{\mathrm{B}}$. The influence of the magnetic
field is most significant for $\rho \leq \rho _{\mathrm{B}}$ and
$T\leq T_{\mathrm{B}}$ when electrons occupy the ground Landau level.
In this case one can deal with the strong quantizied gas and the magnetic
field modifies all parameters of the electron gas. For example, for
degenerate, non-relativistic electrons the pressure is proportional
to $\rho ^{\mathrm{3}}$ and this form of the equation of state which
one can compare with that for the case B=0 for which $P\sim \rho ^{\mathrm{5/3}}$.
When $\rho >\rho _{\mathrm{B}}$ the Fermi temperature is still greater
than the magnetic temperature $T_{\mathrm{B}}$. Electrons are degenerate
and there are many Landau levels, now the level spacing exceeds $k_{\mathrm{B}}T$.
The properties of the electron gas are only slightly affected by the
magnetic field. With increasing temperature, there is the thermal
broadening of Landau levels, when $T\geq T_{\mathrm{B}}$, the free
field results are recovered. For $T\gg T_{\mathrm{B}}$ there are
many Landau levels and the thermal widths of the Landau levels are
higher than the level spacing. The magnetic field does not affects
the thermodynamic properties of the gas (Lai 2000 ).\\
 The total pressure of the system can be described as the sum of the
pressure coming from electrons and ions plus small corrections coming
from the magnetic field \[
P=P_{\mathrm{e}}+P_{\mathrm{QED}}.\]
 The contributions from the same constituents form the energy density
\[
\varepsilon =\varepsilon _{\mathrm{ion}}+\varepsilon _{\mathrm{QED}},\]
 where $P_{\mathrm{QED}}=\frac{1}{6}B^{\mathrm{2}}$, $\varepsilon _{\mathrm{QED}}=\frac{1}{2}B^{\mathrm{2}}$.
The electron pressure is defined with the use of the Fermi integral
\begin{eqnarray}
 & P_{\mathrm{e}}=\frac{2\gamma m_{\mathrm{e}}^{\mathrm{4}}}{4\pi ^{\mathrm{2}}}\sum _{n=0}^{\infty }[2-\delta _{\mathrm{n0}}]\times  & \\
 & I_{\mathrm{-1,2,+}}(z/t,1+2\gamma n)+I_{\mathrm{-1,2,-}}(z/t,1+2\gamma n)). & \nonumber 
\end{eqnarray}
 The ion energy density has the simple form \\
\begin{equation}
\varepsilon _{\mathrm{ion}}=\left(\frac{A}{Z}\right)n_{\mathrm{e}}m_{\mathrm{B}},\end{equation}
 where $m_{\mathrm{B}}$ is the baryon mass. The ions are very heavy
and are treated as classical gas. The influence of the magnetic field
for ions is very small and this correction to the equation of state
has been neglected.\\
 The pressure and energy density dependence on the chemical potential
$\mu _{\star }\, \, $ determines the form of the equation of state,
which is calculated in the flat Minkowski space time. The obtained
form of the equation of state is the fundamental input in determining
macroscopic properties of a star. Outside the static spherically symmetric
body the metric has the form \begin{eqnarray*}
ds^{\mathrm{2}}=e^{\nu \mathrm{(r)}}dt^{\mathrm{2}}-e^{\lambda \mathrm{(r)}}dr^{\mathrm{2}}-r^{\mathrm{2}}d\theta ^{\mathrm{2}}-r^{\mathrm{2}}\sin ^{\mathrm{2}}\theta d\varphi ^{\mathrm{2}}, &  & 
\end{eqnarray*}
 where $\nu (r)$ and $\lambda (r)$ are metric functions. The covariant
components of the metric tensor \begin{equation}
g_{\mu \nu }=\left(\begin{array}{cccc}
 e^{\nu } & 0 & 0 & 0\\
 0 & -e^{\lambda } & 0 & 0\\
 0 & 0 & -r^{\mathrm{2}} & 0\\
 0 & 0 & 0 & -r^{\mathrm{2}}sin^{\mathrm{2}}\theta \end{array}
\right).\label{tensormetryczny2}\end{equation}
 together with the specified form of the energy momentum tensor $T_{\mu \nu }$
allows to derive the structure equations of a spherically symmetric
star (Tolman Oppenheimer Volkoff equations) \begin{eqnarray}
 & \frac{dP(r)}{dr}= & \label{otv}\\
= & -\frac{G}{r^{\mathrm{2}}}(\rho (r)+P(r))\frac{(m(r)+4\pi P(r)r^{\mathrm{3}})}{(1-\frac{2Gm(r)}{r})} & \nonumber 
\end{eqnarray}
\begin{eqnarray}
\frac{dm(r)}{dr}=4\pi r^{\mathrm{2}}\rho (r). &  & \label{teq2}
\end{eqnarray}
 As the solution the mass-radius relation for white dwarfs is obtained.\\
 The analyzed model of the white dwarf has to include both the core
of degenerate isothermal electrons and the atmosphere where the temperature
changes with the radius. In this case the equation of hydrostatic
equilibrium has to be supplemented by the transport equation \begin{eqnarray}
\frac{d(e^{\nu \mathrm{(r)}}T)}{dr}=-\frac{3}{16\sigma }\frac{\kappa \rho }{T_{\star }^{\mathrm{3}}}\frac{L}{4\pi r^{\mathrm{2}}}\frac{e^{\nu \mathrm{(r)}}}{\sqrt{1-\frac{2Gm}{r}}}, &  & \label{lumi}
\end{eqnarray}
 where $e^{\nu \mathrm{(r)}}$ is the redshift, $L$ is the luminosity,
$\kappa $ is the opacity of the stellar material and $\sigma $ is
the Boltzmann constant. After neglecting small corrections coming
from the general relativistic form of the equation (\ref{lumi}) and
using the condition of isothermality $T_{\star }=const$ it is possible
to determine the luminosity of a white dwarf. In the result the total
radius of a white dwarf can be obtained from the relation \begin{eqnarray}
R=\left(\frac{1}{R_{\star }}-\frac{T_{\star }}{GM}\frac{k_{\mathrm{B}}}{\widetilde{\mu }}4.25\right)^{\mathrm{-1}}, &  & 
\end{eqnarray}
 ( Shapiro et al. 1983 ). In this relation $R_{\star }$ is the core
radius and $\widetilde{\mu }$ is the molecular weight of the gas
particles. The boundary conditions for the equation (\ref{otv}) are
giving in the following form: $P(r=0)=P_{\mathrm{c}}=P(\rho _{\mathrm{c}})$
where $\rho _{\mathrm{c}}$ denotes the density at the center of the
star which in this paper ranges from $10^{\mathrm{5}}$ $g$ $cm^{\mathrm{-3}}$
to $10^{\mathrm{11}}$ $g$ $cm^{\mathrm{-3}}$. The condition $P(r=R)=0$
determines the star radius $R$. For such boundary conditions the
value of the magnetic field and temperature are limited to the following
values $T=0-10^{\mathrm{7}}$ $K$, $\gamma =0-0.1$, respectively.
The results are presented in figures 1-3.\\
 Studying the properties of a star in the framework of the general-relativistic
Thomas-Fermi model one can made the assumption that the temperature
and chemical potential are metric dependent local quantities and the
gravitational potential instead of Poisson's equation satisfies Einstain's
field equations. The energy momentum conservation $T_{;\nu }^{\mu \nu }=0$
which for spherically symmetric metric (\ref{tensor metryczny2})
can be written as \begin{eqnarray}
\frac{d\nu }{dr}=-\frac{2}{P+\rho }\frac{dP}{dr} &  & \label{teq}
\end{eqnarray}
 together with the Gibbs-Duhem relation and with the assumption that
the heat flow and diffusion vanishes, (Israel 1976) give the condition
\begin{eqnarray}
\frac{\mu _{\star }}{T_{\star }}=const. &  & \label{eq19}
\end{eqnarray}
 This implies that the temperature and chemical potential become local
metric functions \begin{eqnarray}
T(r) & = & e^{\mathrm{-}\nu \mathrm{(r)/2}}T_{\star }\label{tol1}\\
\mu (r) & = & e^{\mathrm{-}\nu \mathrm{(r)/2}}\mu _{\star }\label{tol2}
\end{eqnarray}
 where $T_{\star }$ and $\mu _{\star }$ are constant. The first
equation in (\ref{tol1}) is the well known Tolman condition for thermal
equilibrium in a gravitational field, (Tolman 1934). Now, the Fermi-Dirac
distribution function has to be redefined and is given in the following
form \begin{eqnarray*}
 & \frac{1}{e^{\left(\sqrt{1+x^{\mathrm{2}}+2\gamma n}-z_{\mathrm{0}}\right)\slash \mathrm{t}_{\star }}+1}\rightarrow \frac{1}{e^{\left(e^{\nu \mathrm{(r)/2}}\sqrt{1+x^{\mathrm{2}}+2\gamma n}-z_{\mathrm{0}}\right)\slash \mathrm{t}_{\star }}+1} & \\
 & =\frac{1}{e^{\left(\sqrt{1+x^{\mathrm{2}}+2\gamma n}-z(r)\right)\slash \mathrm{t(r)}}+1}, & 
\end{eqnarray*}
 ( Bili\'{c} et al. 1999). In the result instead of the OTV equilibrium
equations together with the equation of state which has been calculated
in the flat space-time one can derived three self-consistent equations
(\ref{otv},\ref{teq2},\ref{teq}). These equations together with
the local form of the equation of state are now functions of $r$.
This is of particular importance in the case of strong gravitational
fields.

\section{Discussion}

In order to construct the mass-radius relation for white dwarfs the
proper form of the equation of state have to be enumerated. For our
needs we have chosen the plasma as the main constituent of the theoretical
model of the white dwarf interior. All calculations have been performed
at finite temperature and different from zero magnetic field and compared
with those of zero temperature and without magnetic field.\\
 In Fig. 1 the applied forms of the equations of state for magnetic
and non-magnetic white dwarfs have been presented. Dots indicate the
location of critical points. The approximation of spherical symmetry
can be used for densities grater than the critical one $\rho _{crit}$.
The inclusion of the magnetic field makes the equation of state stiffer.
In the case of zero temperature model there are clearly visible Landau
levels which became smear out when the finite temperature is included.
The figure in the subpanel shows the equation of state obtained for
the low density range for the zero temperature case, Landau levels
are visible. In Fig. 2 the mass-radius relations for magnetic white
dwarfs are presented. For each value of $\gamma $ the maximum mass
has nearly the same value however the approximation of spherical symmetry
which is connected with the existence of critical point determines
the configuration with maximum radius. For the sake of completeness
this figure includes also the mass-radius relation obtained for the
equation of state of Hamada and Salpeter (Hamada et al. 1961). The
solid line represents the nonmagnetic case which has been get with
the use of the Hamada Salpeter equation of state. Dashed and dotted
curves have been constructed for different values of the magnetic
field strength. The differences between these curves are very small.
The most important effect concerns the position of the critical points
connected with the applied approximation of the spherical symmetry.
In this approximation stable white dwarf configurations exist only
for densities higher then $\rho _{crit}$. Thus the critical points
on the curves determine configurations with maximal radius $R$. The
stronger the magnetic field the smaller the white dwarf radius is.
The most extended objects are obtained for the weakest magnetic fields.
Using the Thomas-Fermi approximation one can obtain the result in
which the star interior is no longer isothermal. The temperature profiles
are presented in Fig. 3 for the fixed values of the central density
$\rho _{\mathrm{c}}$. The temperature has been changed but not in
a significant way. This is in agreement with the value of the gravitational
potential which in the case of white dwarfs is much smaller than that
obtained for neutron stars. The higher the central density the more
visible are changes in temperature inside the star. For the central
density $\rho _{\mathrm{c}}=10^{\mathrm{10}}gcm^{\mathrm{-3}}$ the
temperature difference in the center of the star equals $6\, 10^{\mathrm{4}}$
$K$. The magnetic temperatures $T_{\mathrm{B}}$ for different values
of the central density $\rho _{\mathrm{c}}$ and different magnetic
field strengths are presented in Tables 1, 2, 3. It has been found
that $T_{\mathrm{B}}$ increases with the increasing value of the
magnetic field and decreasing for increasing value of the central
density. In Tables 4, 5, 6 the basic parameters of magnetic and non-magnetic
white dwarfs are collected. For the maximum mass configurations the
atmosphere extent drop with the increasing strength of the magnetic
field. The effective temperature rises with the increasing magnetic
field whereas the luminosity is altered insignificantly. All changes
of white dwarf parameters are evident in the area of moderate densities.
In this paper a magnetization ${\mathcal{M}}$ in the medium is not
included. The inclusion of magnetization changes the perpendicular
to the magnetic field component of a pressure and alters the value
of the star radius. In the paper by Felipe it is stated that for positive
magnetization the transversal pressure exerted by the charged particles
in the magnetic field is smaller than the longitudinal one by the
amount $B{\mathcal{M}_{\mathrm{e}}}$ (Felipe et al. 2002). Thus for
$n=0$ (the ground Landau level) $P_{\perp }=0$. The vanishing of
the pressure means that the instability which leads to the gravitational
collapse appear. However, considering the case of finite temperature
it is necessary to take into account a fact that the temperature smearing
out Landau levels causes the appearance of the residual pressure which
supports the star against gravitational collapse. The instability
criterion given by the condition \begin{eqnarray*}
n_{\mathrm{max}}\simeq 2\, 10^{\mathrm{-21}}\frac{n_{\mathrm{e}}^{\mathrm{2}}}{B_{\mathrm{z}}^{\mathrm{3}}} &  & 
\end{eqnarray*}
 sets in for densities equal $5.7\times 10^{\mathrm{3}}$ $g\, \, cm^{\mathrm{-3}}$
for $\gamma =0.01$, $5\times 10^{\mathrm{4}}$ $g\, \, cm^{\mathrm{-3}}$
for $\gamma =0.05$ and $1.5\times 10^{\mathrm{5}}$ for $\gamma =0.1$.
This instability leads to a cigar like object whereas in the case
considered in this paper gives in the limiting case a toroidal star.
\begin{figure}
\includegraphics[  width=12cm]{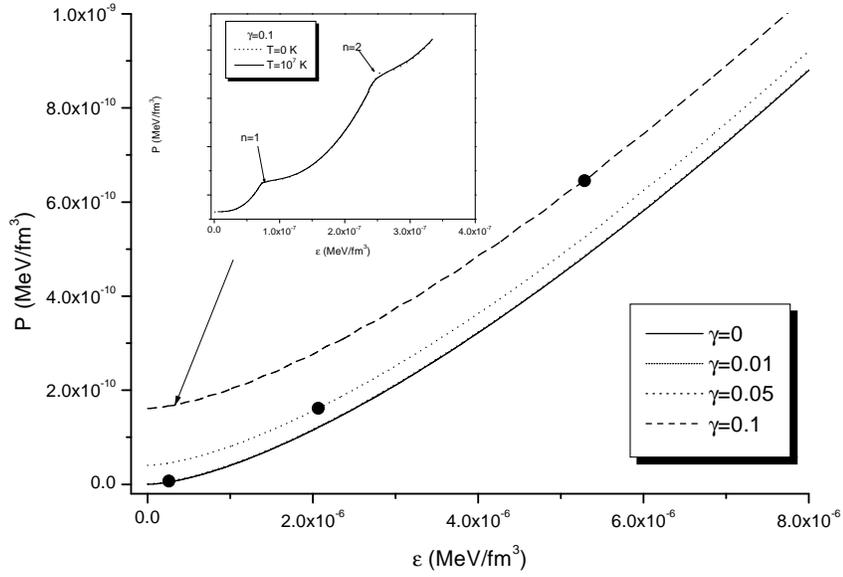}

\caption{The equation of state for different cases of temperature and magnetic
field strength. Dots represent the position of critical points. The
figure in the subpanel depicts the form of the equation of state constructed
for the low density range with clearly visible Landau levels.}
\end{figure}

\begin{figure}
\includegraphics[  width=12cm]{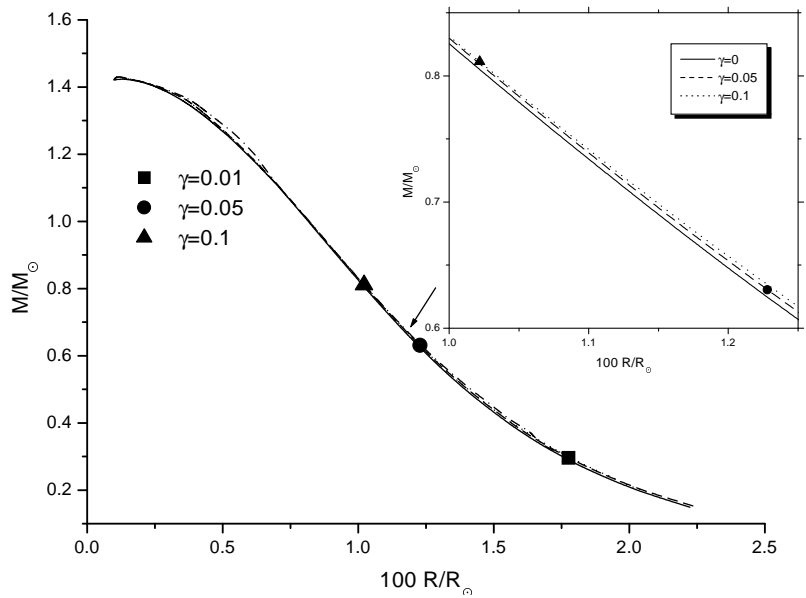}

\caption{The mass-radius relations for magnetic white dwarfs obtained for
different values of the magnetic field strength. The marked points
represent the limiting values of radii for stable spherically symmetric
configurations. The figure in the subpanel shows the mass-radius relations
obtained for the radius range $1-1.25\, R_{\odot }/100$. For comparison
the nonmagnetic case is constructed for the Hamada-Salpeter equation
of state (Hamada et al. 1961).}
\end{figure}

\begin{figure}
\includegraphics[  width=12cm]{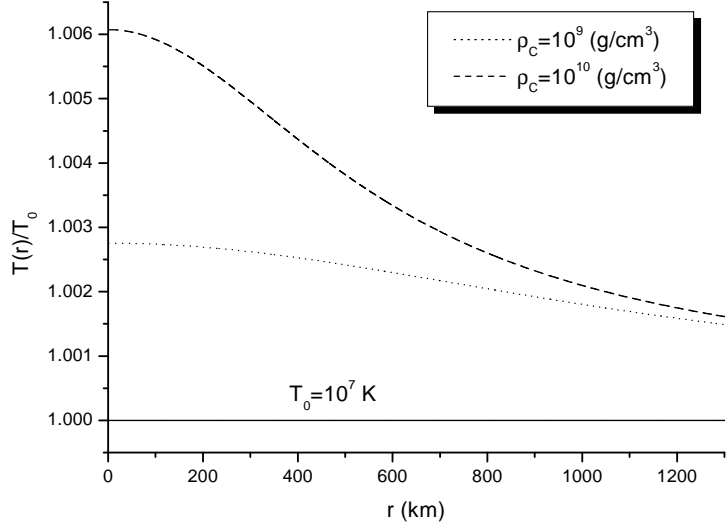}

\caption{The temperature profile for different central densities.}
\end{figure}

\begin{table}
\begin{tabular}{cc}
\hline 
$\rho _{\mathrm{c}}\, [g\, \, cm^{\mathrm{-3}}]$&
 $T_{\mathrm{B}}\, [K]$\\
\hline
$7\, 10^{\mathrm{6}}$&
 $3.2398\, 10^{\mathrm{7}}$\\
\hline
$3\, 10^{\mathrm{7}}$&
 $2.2115\, 10^{\mathrm{7}}$\\
\hline
$3.5\, 10^{\mathrm{10}}$&
 $2.2631\, 10^{\mathrm{6}}$ \\
\hline
\end{tabular}

\caption{The magnetic temperature $T_{\mathrm{B}}$ for the white dwarf with
$\gamma =0.01$ and temperature $T=0$. }

\label{tab: a}
\end{table}
 \newpage

\begin{table}
\begin{tabular}{cc}
\hline 
$\rho _{\mathrm{c}}\, [g\, \, cm^{\mathrm{-3}}]$&
 $T_{\mathrm{B}}\, [K]$\\
\hline
$7\, 10^{\mathrm{6}}$&
 $1.6199\, 10^{\mathrm{8}}$\\
\hline
$3\, 10^{\mathrm{7}}$&
 $1.1088\, 10^{\mathrm{8}}$\\
\hline
$3.5\, 10^{\mathrm{10}}$&
 $1.1315\, 10^{\mathrm{7}}$ \\
\hline
\end{tabular}

\caption{The magnetic temperature $T_{\mathrm{B}}$ for the white dwarf with
$\gamma =0.05$ and temperature $T=0$. }

\label{tab: b}
\end{table}

\begin{table}
\begin{tabular}{cc}
\hline 
$\rho _{\mathrm{c}}\, [g\, \, cm^{\mathrm{-3}}]$&
 $T_{\mathrm{B}}\, [K]$\\
\hline
$7\, 10^{\mathrm{6}}$&
 $3.2647\, 10^{\mathrm{8}}$\\
\hline
$3\, 10^{\mathrm{7}}$&
 $2.2255\, 10^{\mathrm{8}}$\\
\hline
$3.5\, 10^{\mathrm{10}}$&
 $2.2632\, 10^{\mathrm{7}}$ \\
\hline
\end{tabular}

\caption{The magnetic temperature $T_{\mathrm{B}}$ for the white dwarf with
$\gamma =0.1$ and temperature $T=0$. }

\label{tab: c}
\end{table}

\begin{table}
\begin{tabular}{cccccc}
\hline 
$1$&
 $2$&
 $3$&
 $4$&
 $5$&
 $6$\\
\hline
$1.083$&
 $1.099$&
 $0.0158$&
 $13679$&
 $-$&
 $0$\\
\hline
$1.089$&
 $1.105$&
 $0.0159$&
 $15822$&
 $4.6\, 10^{\mathrm{5}}$&
 $0.01$\\
\hline
$1.087$&
 $1.103$&
 $0.0154$&
 $15896$&
 $3.6\, 10^{\mathrm{6}}$&
 $0.05$\\
\hline
$-$&
 $-$&
 $-$&
 $-$&
 $-$&
 $0.1$ \\
\hline
\end{tabular}

\caption{The parameters of the magnetic and non-magnetic white dwarf with
the mass $0.74$ $M_{\odot }$, central density $\rho _{\mathrm{c}}=7\, 10^{\mathrm{6}}g\, \, cm^{\mathrm{-3}}$
and luminosity $L=0.0012$ $L_{\odot }$. }

\label{tab: 1}
\end{table}
 \newpage

\begin{table}
\begin{tabular}{cccccc}
\hline 
$1$&
 $2$&
 $3$&
 $4$&
 $5$&
 $6$\\
\hline
$0.807$&
 $0.815$&
 $0.00870$&
 $24633$&
 $-$&
 $0$\\
\hline
$0.810$&
 $0.818$&
 $0.00877$&
 $30799$&
 $4.6\, 10^{\mathrm{5}}$&
 $0.01$\\
\hline
$0.809$&
 $0.818$&
 $0.00875$&
 $30864$&
 $3.6\, 10^{\mathrm{6}}$&
 $0.05$\\
\hline
$0.807$&
 $0.816$&
 $0.00872$&
 $30991$&
 $9.4\, 10^{\mathrm{6}}$&
 $0.1$ \\
\hline
\end{tabular}

\caption{The parameters of the magnetic and non-magnetic white dwarf with
the mass $1$ $M_{\odot }$, central density $\rho _{\mathrm{c}}=3\, 10^{\mathrm{7}}g\, \, cm^{\mathrm{-3}}$
and luminosity $L=0.0016$ $L_{\odot }$. }

\label{tab: 2}
\end{table}
\begin{table}
\begin{tabular}{cccccc}
\hline 
$1$&
 $2$&
 $3$&
 $4$&
 $5$&
 $6$\\
\hline
$0.130$&
 $0.130$&
 $0.00022$&
 $940946$&
 $-$&
 $0$\\
\hline
$0.117$&
 $0.117$&
 $0.00018$&
 $1.60\, 10^{\mathrm{6}}$&
 $4.6\, 10^{\mathrm{5}}$&
 $0.01$\\
\hline
$0.117$&
 $0.117$&
 $0.00018$&
 $1.61\, 10^{\mathrm{6}}$&
 $3.6\, 10^{\mathrm{6}}$&
 $0.05$\\
\hline
$0.111$&
 $0.111$&
 $0.00016$&
 $1.78\, 10^{\mathrm{6}}$&
 $9.4\, 10^{\mathrm{6}}$&
 $0.1$ \\
\hline
\end{tabular}

\caption{The parameters of the magnetic and non-magnetic white dwarf with
the mass $1.42$ $M_{\odot }$, central density $\rho _{\mathrm{c}}=3.5\, 10^{\mathrm{10}}g\, \, cm^{\mathrm{-3}}$
and luminosity $L=0.0023$ $L_{\odot }$. }

\label{tab: 3}
\end{table}
The numbers from 1 to 6 denote:\\
 1-the core radius of the white dwarf $100R_{\star }/R_{\odot }$\\
 2-the total radius of the white dwarf $100R/R_{\odot }$\\
 3-the extent of the atmosphere $100\Delta R/R_{\odot }$\\
 4-the effective temperature $T_{\mathrm{eff}}$ $[K]$,\\
 5-the critical density $\rho _{\mathrm{crit}}$ $[g\, \, cm^{\mathrm{-3}}]$,\\
 6-the strength of magnetic field $\gamma $

\end{document}